\documentclass[runningheads]{llncs}

\usepackage[T1]{fontenc}
\usepackage{graphicx}
\usepackage{xurl} 
\usepackage{amsmath}
\usepackage[utf8]{inputenc}
\usepackage{subfigure}
\usepackage{gensymb}
\usepackage{array}
\usepackage{booktabs}
\usepackage{tabularx,booktabs}
\usepackage{cite}
\usepackage{color}
\usepackage{alltt}
\usepackage[hidelinks]{hyperref}
\usepackage{enumerate}
\usepackage{siunitx}
\usepackage{epstopdf}
\usepackage{pbox}
\usepackage{bbding}
\usepackage[linesnumbered, ruled, vlined, noend]{algorithm2e}
\usepackage{diagbox}
\usepackage{subcaption}
\usepackage{multirow}

\usepackage[flushleft]{threeparttable}
\usepackage{float}
\usepackage{comment}
\usepackage{adjustbox}
\usepackage{makecell}
\usepackage{amssymb}

\newcolumntype{C}{>{\centering\arraybackslash}X} 
\begin{document}

\title{TYTAN: Taylor-series based Non-Linear Activation Engine for Deep Learning Accelerators}

\titlerunning{TYTAN: Taylor-series based Non-Linear Activation Engine}

\author{Soham Pramanik\inst{1}\orcidID{0009-0005-1772-2901} \and
Vimal William\inst{2}\orcidID{0000-0002-1069-9594} \and
Arnab Raha\inst{3}\orcidID{0000-0002-8848-1069} \and
Debayan Das\inst{4}\orcidID{0000-0003-1843-0124} \and
Amitava Mukherjee\inst{5,6}\orcidID{0000-0003-1694-3911} \and
Janet L. Paluh\inst{6}\orcidID{0000-0002-5988-6075}}


\authorrunning{S. Pramanik et al.}

\institute{Department of Electronics and Telecommunication Engineering, Jadavpur University, Kolkata, India \and
SandLogic Technologies, Bangalore, India \and
Intel Corporation, Santa Clara, CA, USA \and
Department of Electronic Systems Engineering, Indian Institute of Science, Bangalore, India \and
Department of Computer Science and Engineering, Amrita University, Amritapuri, Kollam, Kerala, India \and
College of Nanoscale Science and Engineering, Nanobioscience, SUNY Polytechnic Institute, Albany, New York, USA}

\maketitle

\begin{abstract}
The rapid advancement in AI architectures and the proliferation of AI-enabled systems have intensified the need for domain-specific architectures that enhance both the acceleration and energy efficiency of AI inference, particularly at the edge. This need arises from the significant resource constraints—such as computational cost and energy consumption—associated with deploying AI algorithms, which involve intensive mathematical operations across multiple layers. High-power-consuming operations, including General Matrix Multiplications (GEMMs) and activation functions, can be optimized to address these challenges. Optimization strategies for AI at the edge include algorithmic approaches like quantization and pruning, as well as hardware methodologies such as domain-specific accelerators. This paper proposes TYTAN: TaYlor-series based non-linear acTivAtion eNgine, which explores the development of a Generalized Non-linear Approximation Engine (G-NAE). TYTAN targets the acceleration of non-linear activation functions while minimizing power consumption. The TYTAN integrates a re-configurable hardware design with a specialized algorithm that dynamically estimates the necessary approximation for each activation function, aimed at achieving minimal deviation from baseline accuracy. The proposed system is validated through performance evaluations with state-of-the-art AI architectures, including Convolutional Neural Networks (CNNs) and Transformers. Results from system-level simulations using Silvaco's FreePDK45 process node demonstrate TYTAN's capability to operate at a clock frequency $>950 MHz$, showcasing its effectiveness in supporting accelerated, energy-efficient AI inference at the edge, which is $\sim2\times$ performance improvement, with $\sim56\%$ power reduction and $\sim35\times$ lower area compared to the baseline open-source NVIDIA Deep Learning Accelerator (NVDLA) implementation.

\end{abstract}

\section{Introduction}

The continual advancements in Artificial Intelligence (AI) and the evolving demands for AI-assisted applications highlighted the limitations of existing infrastructures on processing such as data-intensive operations. These limitations significantly elevated the cost of accommodating modern-day AI-based applications and upgrading the existing infrastructure to support such applications could also be challenging. With the motivation of proliferating the benefits of AI, this paper focuses on optimizing the effective computation of non-linear activation functions employed in Neural Networks (NN) through state-of-the-art hardware-aware DNN algorithm optimization and algorithm-aware hardware design. The method aimed to improve the inference speed by accelerating the above-mentioned compute functions in the AI model. In addition, the methods focused on extending the support for AI-based solutions in resource-constrained environments.

\subsection{Background}

The rapid evolution of AI workloads demands optimized hardware infrastructures for modern deep learning models. A key bottleneck in AI accelerators is computing nonlinear functions like \textit{softmax}, \textit{sigmoid}, \textit{GeLU}, and \textit{layer normalization}. While multiply-accumulate (MAC) operations are well-optimized, efficient nonlinear operations implementation remains challenging.

Prior works address these challenges through approximate computation techniques. \textit{ViTALiTy} \cite{10071081} approximates \textit{softmax} in linear attention layers using Taylor series expansion. \textit{NN-LUT} \cite{10.1145/3489517.3530505} employs look-up-table (LUT)-based approximation for nonlinear functions, optimizing transformer models with precomputed values. However, LUT methods, while low-latency, suffer from high memory costs and limited scalability. 

Alternative approximation techniques include \textit{Fast Approximations of Activation Functions} \cite{fastapprox2021posit} using \textit{Posit arithmetic}, \textit{TaylorNet} \cite{taylornet2020approx} with adaptive Taylor expansions reducing computational cost by up to 60\%, and comparative studies of LUT-based, CORDIC, and Taylor series methods \cite{approxactivation2022}.

\textit{UNO} \cite{9502473,wu2021uno} introduces a MAC-based architecture unifying nonlinear function computation by mapping Taylor-approximated functions onto existing SIMD-based MAC arrays, eliminating dedicated processing units and enabling dynamic accuracy-energy scaling. However, UNO primarily uses low-order Taylor approximations, potentially compromising accuracy in complex workloads.

Building upon these advancements, this work integrates both algorithmic and hardware design optimizations to enhance nonlinear function mapping in AI accelerators. TYTAN achieves a better balance between accuracy, power, and area efficiency, ensuring minimal resource overhead while maintaining better performance and scalability than past attempts.

\subsection{Motivation}
    A generalized and re-configurable activation engine for accelerating activation functions with minimal power consumption can be a promising method to handle the challenges addressed previously. The co-design of approximated computing and generalized hardware architecture for accelerating non-linear activation functions shows greater potential to enable ultra-low-powered AI inference at the edge. The proposed system combines re-configurable hardware design with a specialized algorithm for estimating the amount of approximation required per activation function, providing reduced power consumption with minimal deviation from the AI model's baseline accuracy.
    
    The mathematical limitations of the Taylor series toward the representation of discontinuous functions restrict the proposed hardware design from supporting piece-wise linear functions such as ReLU. However, ReLU requires only trivial hardware implementation (simple sign bit check), making it unsuitable for acceleration. TYTAN targets computationally intensive non-linear functions that involve expensive operations where hardware acceleration provides substantial benefits. The key contributions of this paper are summarized below:
    
    \begin{itemize}
      \item We propose a Taylor-series-based Non-Linear Activation Engine, TYTAN to accelerate a wide range of approximated non-linear activation functions. 
      \item An iterative search-based algorithm is designed to estimate the amount of approximation required per activation function based on the layer placement and correlation with the AI model's baseline accuracy.
      \item The proposed GNAE system design combines the TYTAN accelerator with add-ons to support a wide range of non-linear functions providing a modular and scalable solution. 
      \item Using Silvaco's FreePDK45 process node \cite{opencell_freepdk}, we demonstrate a hardware accelerator for various non-linear activation functions operating at a clock frequency of $950MHz$, showing $>56\%$ power reduction, $\sim 2\times$ performance improvement, and $>35\times$ lower area compared to the baseline open-source NVIDIA Deep Learning Accelerator NVDLA \cite{NVDLA} implementation.
    \end{itemize}

\section{Methodology}
    The proposed architecture articulated in Fig. 1 is a software-hardware co-design for accelerating non-linear activation functions utilized in modern AI architectures. The algorithm is tailored to intake AI models from multiple frameworks and is dedicated to computing the approximation required by individual activation functions. Furthermore, the hardware can be programmed with the pre-calculated coefficients into its buffer to expedite the computing of the appropriate nonlinear functions. The detailed infrastructure of software and hardware sub-systems utilized in the TYTAN is described in the following sections.

    \begin{figure}[t]
        \includegraphics[width=\linewidth]{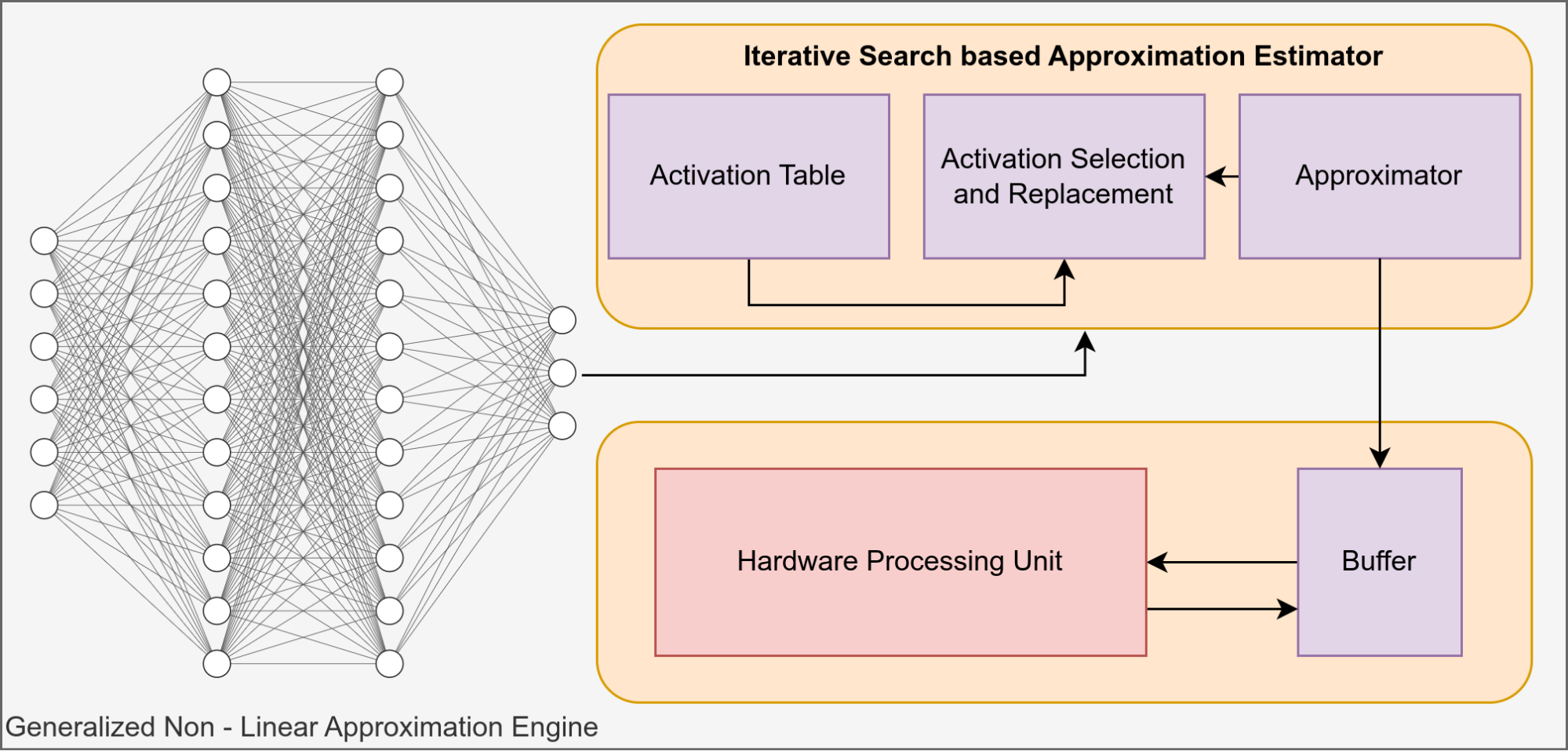}
        \caption{ Software/Hardware Co-design for Generalized Non-linear Approximation Engine}
        \label{intro}
    \end{figure}

\subsection{Algorithm Development}
    The software component of the co-design includes a well-structured, iterative search algorithm that determines the optimal number of terms, n, required in the Taylor series expansion to approximate the non-linear functions used in AI architectures. Additionally, the algorithm incorporates a simulated version of each approximated non-linear function, as described in Eqs. \ref{eq:4}, \ref{eq:5}, \ref{eq:6}, \ref{eq:7}, and \ref{eq:8}, and replaces the original functions in the AI model with these approximations. Furthermore, the algorithm dynamically computes the optimal number of terms, n, based on the function’s position within the model and a predefined accuracy budget, ensuring minimal degradation in baseline accuracy. The approximator described in Fig. \ref{intro} includes the above-illustrated search algorithm for identifying non-linear activation functions in the AI architecture. The activation selection and replacement process involves replacing the selected activation function with the pre-defined approximated activation from the activation table. The approximator triggers a counter at every activation selection and replacement, which computes the order of the Taylor series with the pre-defined search range. The search space for the approximation is determined through a set of brute-force experiments to understand the point of convergence of the approximated functions with the standard functions. The approximator also maintained a predefined deviation from the baseline accuracy by comparing the baseline accuracy to the estimated accuracy of the evaluated neural network at each iteration in computing the order of the Taylor series. 
    
    The algorithm is generic and can support various other non-linear activation functions similar to the above-described functions. The number of terms $n$ in the approximated activation function of the Taylor series is directly proportional to the hardware resource requirements. The search space for approximation is constrained to balance power consumption and deviation from the AI model’s baseline accuracy. Algorithm \ref{alg:approximator} presents the detailed infrastructure of the aforementioned algorithm used in TYTAN. The estimated approximation by the algorithm can be determined by factors such as activation placement in the AI architecture and its impact on the feature map to its preceding layers in the AI model. The approximation estimation is dynamic, and the depth of the Taylor series varies depending on the individual activation function based on the factors mentioned above. Utilizing pre-trained AI models as input can significantly lower the run-time cost of the algorithm, and it also inputs a minimal amount of dataset for accuracy estimation at every iteration. The algorithm's run-time complexity depends on the complexity of the input AI architecture. The framework supports retraining AI models with the approximated non-linear activation functions by its pluggable infrastructure.

\begin{algorithm}[h]
    \caption{General Purpose Non-linear Approximation Algorithm}
    \label{alg:approximator}
    \KwIn{Neural Network Model (\textit{NN Model}), Test Data, Acceptable Accuracy Deviation (\textit{Deviation})}
    \KwOut{Optimized Neural Network Model (\textit{ApproxModel})}
    
    $L \gets \text{ActivationToBeApprox}(\textit{NN Model})$\;
    $BAcc \gets \text{Evaluate}(\textit{NN Model})$\;
    $ModelData \gets [\;]$\;
    
    \ForEach{$Layer \in L$}{
        $LayerData \gets \text{getLayer}(\textit{NN Model}, Layer)$\;
        $[nTerms, Acc] \gets \text{IterativeSearchBasedApprox}(\textit{NN Model}, \textit{Test Data})$\;
        $ModelData.\text{append}([nTerms, Acc])$\;
        
        \If{$BAcc - Acc > \textit{Deviation}$}{
            \textbf{break}\;
        }
    }
    
    $ApproxModel \gets \text{Approximate}(\textit{ModelData}, \textit{NN Model})$\;
    $DeltaAcc \gets BAcc - \text{Evaluate}(\textit{ApproxModel})$\;
    
    \If{$\textit{DeltaAcc} > \textit{Deviation}$}{
        \textbf{call} \textit{Approximator}(\textit{ApproxModel}, \textit{Test Data}, \textit{Deviation})\;
    }
    
    \Return{$ApproxModel$}\;

\end{algorithm}

The above-discussed algorithm focused on improving the hardware acceleration for approximated non-linear activation function through dynamic approximation of every non-linear activation function utilized in the considered AI architecture. The algorithm can be scaled by considering hardware-aware factors, including support for handling higher depths of the Taylor series for better convergence with minimal utilization of system resources. Additionally, the dynamic lower and upper limits of the algorithms for Taylor-series computation enable it to work under different resource constraints, especially with limited hardware resources.

\subsection{Hardware Modeling}
    The proposed hardware design, illustrated in Fig. \ref{intro}, implements an optimal hardware/software co-design strategy to accelerate approximated non-linear activation functions in modern AI architectures. The architecture is reconfigurable and dynamic, enabling efficient power consumption and improved execution speed. The element-wise operator within the non-linear activation functions is mapped using TYTAN, which compute the approximated activation of the incoming tensor. The number of coefficients stored in the internal FIFO buffer determines the accuracy of approximation. The MAC units are modified from conventional MAC designs to implement the algorithm, as expressed in Eq. \ref{eq:3}. 
    
    To establish the mathematical foundation of this approach, the {$e^x$} Taylor series expansion is considered:
    
    \begin{equation} 
        \frac{x^n}{n!} = 1 + x + \frac{x^2}{2!} + \frac{x^3}{3!} + \frac{x^4}{4!} + \dots \label{eq:1}
    \end{equation}
    
    For computational efficiency, this expansion is restructured as follows:
    
    \begin{equation}
        = 1 + x + \frac{x^2}{2!} + x^3 \left[ c_3 + c_4(x) \right] \label{eq:2}
    \end{equation}
    
    This reformulation $T(x)$ enables efficient implementation in TYTAN using a nested multiplication approach, significantly reducing computational complexity:
    
    \begin{equation}
        \textit{T}(x) = \left[ c_0 + x \left[ c_1 + x \left[ c_2 + x \left[ c_3 + c_4(x)
        \right] \right] \right] \right] \label{eq:3}
    \end{equation}
    
    \begin{figure}[!t]
        \centering
        \includegraphics[width=0.9\textwidth]{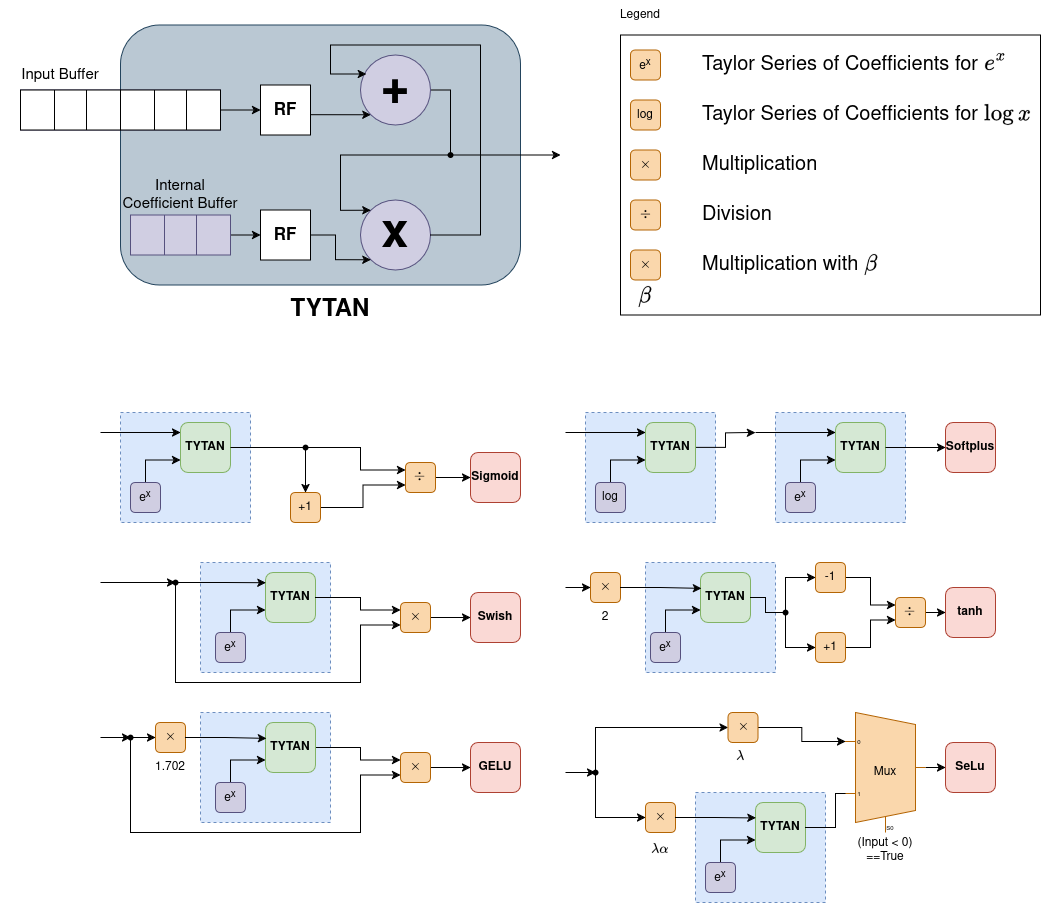}
        \caption{TYTAN Architecture and Example Modes of Operations\vspace{-0.2in}}
        \label{fig:Panel}
    \end{figure}
    
    TYTAN processes four-dimensional tensor inputs as scalar batches using precomputed Taylor series coefficients derived from the algorithm for efficient non-linear function computation. The Taylor series terms ($n$) are precomputed by Algorithm \ref{alg:approximator} and stored in internal hardware buffers to control computational overhead. These variable terms enable flexible, adaptable architecture for efficiently mapping diverse non-linear functions.
    
    The proposed approach can approximate various activation functions commonly used in neural networks. The standard forms of these functions are provided below:
    
    \begin{equation}
        \text{SELU}(x) = \lambda 
        \begin{cases}
            x, & \text{if } x > 0 \\
            \alpha(e^x - 1), & \text{if } x \leq 0
        \end{cases}
    \end{equation}
    
    \begin{equation}
        \text{sigmoid}(x) = \frac{1}{1 + e^{-x}}
    \end{equation}
    
    \begin{equation}
        \text{Swish}(x) = x \cdot \text{sigmoid}(x) = \frac{x}{1 + e^{-x}}
    \end{equation}
    
    \begin{equation}
        \text{GELU}(x) \approx x \cdot \sigma(1.702x) = \frac{x}{1 + e^{-1.702x}}
    \end{equation}
    
    \begin{equation}
        \tanh(x) = \frac{e^x - e^{-x}}{e^x + e^{-x}} = \frac{e^{2x} - 1}{e^{2x} + 1}
    \end{equation}
    
    \begin{equation}
        \text{Softplus}(x) = \log(1 + e^x)
    \end{equation}
    
    The key innovation in this approach lies in efficiently mapping these complex activation functions using the TYTAN hardware output. Equations \ref{eq:4} to \ref{eq:9} present the final formulations of the approximated non-linear functions to leverage hardware acceleration:
    
    \begin{equation}
        \text{SELU}(x) =
        \begin{cases} 
            \lambda x, & \text{if } x > 0 \\
            \lambda \cdot \alpha \cdot T_{e^x}(x),-1 & \text{if } x < 0
        \end{cases}
        \label{eq:4}
    \end{equation}
    
    \begin{equation} 
       \text{sigmoid}(x) = \frac{T_{e^x}(x)}{T_{e^x}(x) + 1}
       \label{eq:5}
    \end{equation}
    
    \begin{equation}
        \text{Swish}(x) = x \cdot T_{e^x}(x)
        \label{eq:6}
    \end{equation}
    
    \begin{equation}
        \text{GELU}(x) = x \cdot T_{e^x}(1.702x)
        \label{eq:7}
    \end{equation}
    
    \begin{equation}
        \text{tanh}(x) = \frac{T_{e^x}(2x) - 1}{T_{e^x}(2x) + 1}
        \label{eq:8}
    \end{equation}
    
    \begin{equation}
        \text{Softplus}(x) = T_{log}(T_{e^x}(x))
        \label{eq:9}
    \end{equation}
    
    The subscripts in each equation indicate the specific Taylor series coefficients utilized for each computation type. These reformulated equations are analogous to the six example TYTAN modes shown in Fig. \ref{fig:Panel}
    
    The proposed implementation offers significant flexibility, as any non-linear activation function can be efficiently mapped onto TYTAN by selecting appropriate coefficients. The arrangement of these activation functions within AI models determines the optimal number of terms $n$ to be encoded into the hardware, dynamically adjusting based on accuracy requirements and power constraints. This adaptability allows the system to maintain high efficiency with minimal energy consumption while achieving acceptable accuracy in AI applications. Furthermore, the architecture provides a unified approach for accelerating diverse activation functions without requiring dedicated hardware for each function type.

\section{HW-SW Co-Design \& Results}
    The robust combination of domain-specific hardware and algorithm results in accelerated computation for non-linear activation functions involved in modern AI Algorithms. The hardware design proposed in section II was evaluated and compared with TensorFlow's activation functions for correctness and accuracy. Detailed remarks on the algorithm's performance and the hardware design are discussed in the following subsections.
    
    \begin{table*}[!h]
        \centering
        \captionsetup{justification=centering} \caption{Performance Analysis of the Approximation Algorithm on the MobileViT Model}
        \label{table:ViT}
        \begin{tabular}{lccccccc}
            \toprule
            \textbf{Model} & \textbf{Target} & \begin{tabular}[t]{@{}c@{}}\textbf{Baseline} \\ \textbf{Acc. (\%)}\end{tabular} & \begin{tabular}[t]{@{}c@{}}\textbf{Min.} \\ \textbf{Length}\end{tabular} & \begin{tabular}[t]{@{}c@{}}\textbf{Max.} \\ \textbf{Length}\end{tabular} & \begin{tabular}[t]{@{}c@{}}\textbf{Avg.} \\ \textbf{Length}\end{tabular} & \begin{tabular}[t]{@{}c@{}}\textbf{Est.} \\ \textbf{Acc. (\%)}\end{tabular} & \textbf{Deviation} \\
            \midrule
            \multirow{3}{*}{MobileViT} & \multirow{3}{*}{Swish} & \multirow{3}{*}{83.38} & 07 & 09 & 07.15 & 82.23 & 0.010 \\
                                       &                        &                        & 07 & 19 & 08.81 & 82.95 & 0.005 \\
                                       &                        &                        & 07 & 25 & 14.44 & 83.14 & 0.0025 \\
            \bottomrule
        \end{tabular}
    \end{table*}

\subsection{Software Setup}
    The experiment to evaluate the impact of the proposed algorithm for dynamic approximation utilizing the Taylor series is carried out by implementing approximated activation functions in the TensorFlow library \cite{abadi2016tensorflow}. The TensorFlow simulations were used to compute the length of the Taylor series required by each approximated function to converge with the standard activation function. The approximation algorithm uses the point of convergence to restrict the search space and avoid unwanted computations beyond the convergence point. Further, the experiment focused on approximating the MobileViT \cite{mehta2022mobilevit}, a state-of-the-art neural network with numerous compute-intensive activation functions. The approximation algorithm setup also includes T4 GPUs from Nvidia for extended hardware support. The algorithm utilizes the GPU acceleration to estimate the layer-wise length of the Taylor series for every activation function in the targeted neural network. MobileVit was initially trained on TFflower \cite{tfflowers} with five classes and a training rate of 30 epochs.
    
    The parameters, such as pre-trained MobileViT and a segment of test data, can be passed to the approximation algorithm for the layer-wise dynamic approximation. The algorithm begins with the baseline accuracy estimation for the targeted neural network, followed by the dynamic approximation for the targeted non-linear activation function. A search for the nominal length of the Taylor series is initiated by the algorithm on account of identifying the targeted non-linear activation function. The iterative search starts from the point of convergence to the lower limit, and the nominal length of the Taylor series is estimated for that particular activation function.
    
    Table \ref{table:ViT} demonstrates a clear correlation between the allowable deviation from the baseline accuracy and the length of the Taylor series utilized for approximation. The approximation algorithm operates for multiple deviation amounts from the baseline accuracy of the targeted neural network, and the length of the Taylor series for approximation is estimated to satisfy the constraints of the deviation from the baseline accuracy.
    
    \begin{figure}[t]
        \centering
        \includegraphics[width=0.8\linewidth]{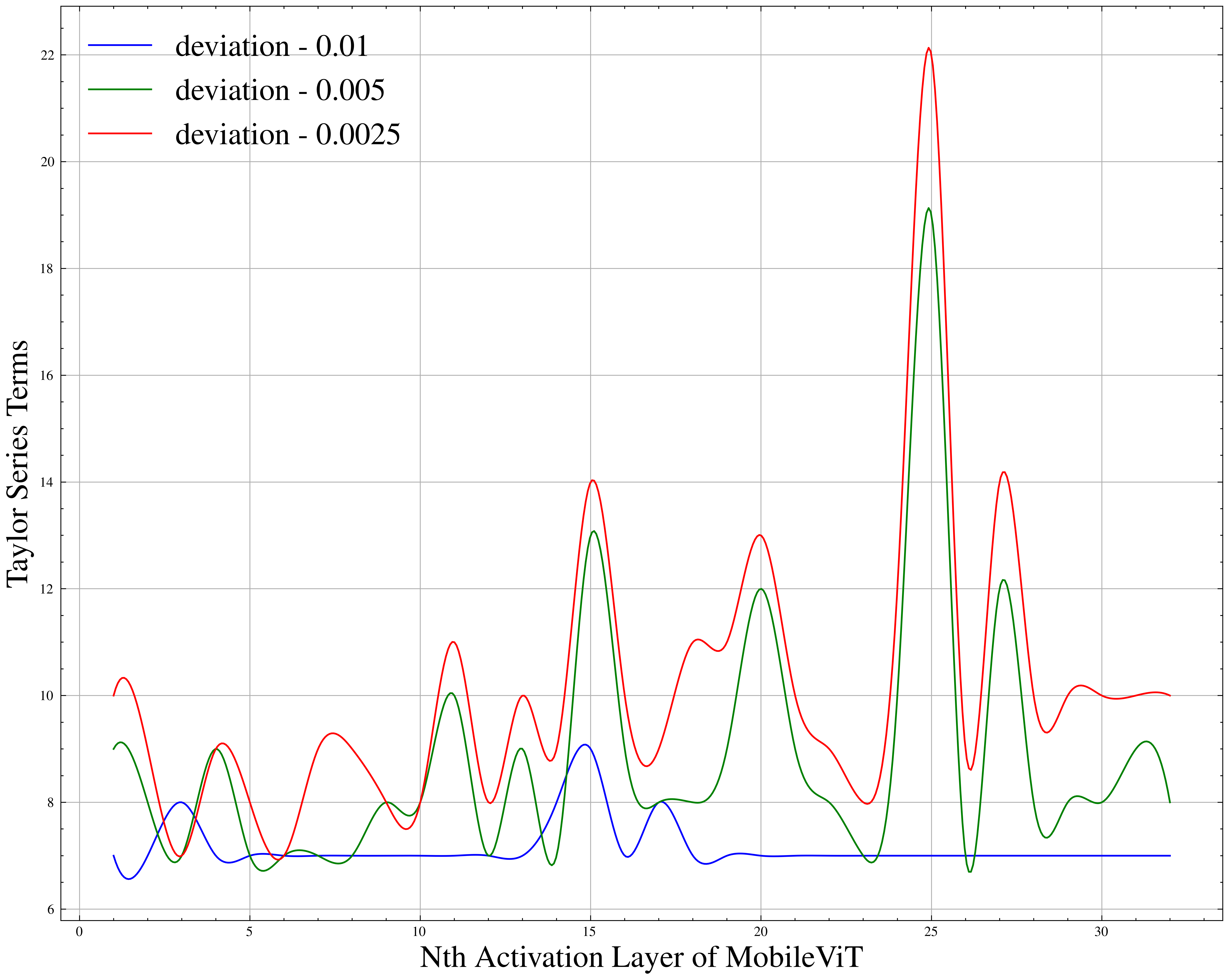}
        \captionsetup{justification=centering} \caption{Comparison of Nth Non-Linear Activation function vs. Order of the Taylor-series per Activation Layer (Swish) of the MobileViT}
        \label{fig:MobileVit}
    \end{figure}

    The algorithm targets about 32 swish non-linear activation functions at the MobileViT. The selection and replacement block described in Fig. \ref{intro} replaces the swish with the approximated swish for every function. Fig. \ref{fig:MobileVit} outlines the required length of the Taylor series for approximation at each targeted activation function at MobileViT for the different amounts of deviations from the baseline accuracy of the targeted neural network. Furthermore, Fig. \ref{fig:MobileVit} also illustrates that the intermediate layers of the considered model are sensitive and require significantly higher Taylor series terms to satisfy the accuracy constraints. It shows that the required length of the Taylor series is minimal for the higher deviation of estimated accuracy from the baseline accuracy of the target neural network and vice versa. The experiment also shows a tight requirement for compute power to elaborate the algorithm for neural networks with more non-linear activation functions. Similarly, the algorithm above and the experiment can be applied to next-generation state-space models like Mamba \cite{gu2024mamba} to enable faster execution of state-of-the-art non-linear functions (Softplus) used within them.

\subsection{Hardware Setup}
    This section provides a detailed elaboration of the hardware realization of the proposed model.

    \begin{figure}[H]
        \centering \includegraphics[width=0.75\linewidth]{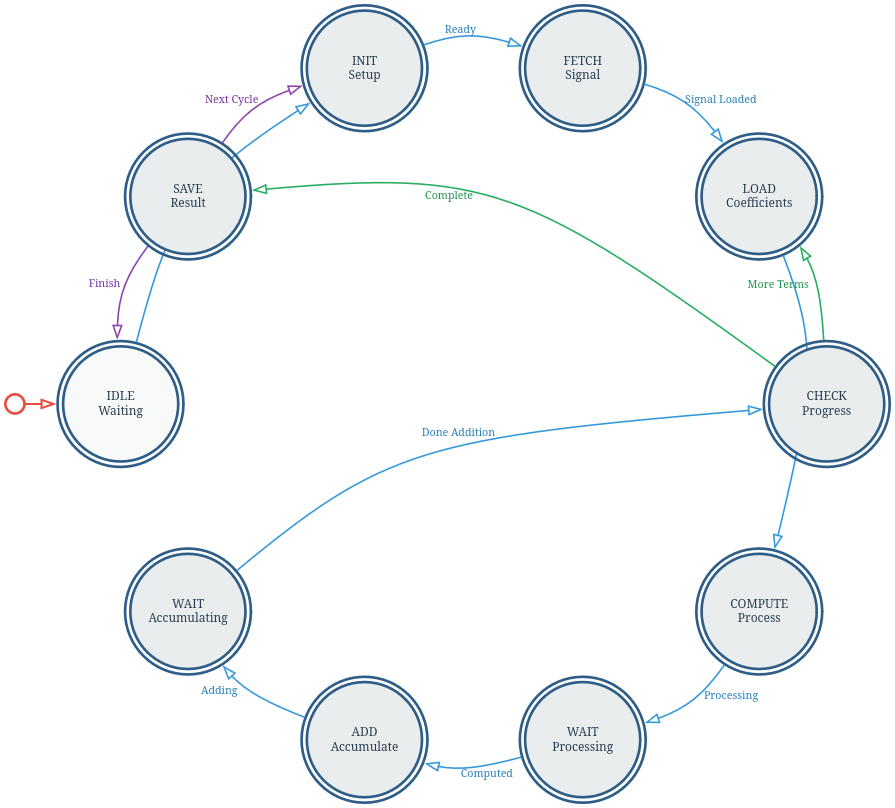}
        \captionsetup{justification=centering} \caption{Finite State Machine (FSM) diagram illustrating the operational flow of the TYTAN architecture}
        \label{fig:fsm}
    \end{figure}

    \begin{table}[t]
        \captionsetup{justification=centering} \caption{TYTAN tanh computation latency with 30 Taylor coefficients}
        \begin{center}
            \begin{tabular}{l>{\centering}p{2.5cm}>{\centering\arraybackslash}p{2.5cm}}
                \toprule
                \textbf{Operation} & \textbf{Clock Cycles Required} & \textbf{Execution Time (ms)\textsuperscript{*}} \\
                \midrule
                Fill buffers (30 input values) & 120 & 0.126 \\
                \addlinespace
                \addlinespace
                Generate output for every input & 747 & 0.786 \\
                \addlinespace
                \addlinespace
                Full TYTAN operation (without Buffers) & 22{,}474 & 23.656 \\
                \addlinespace
                Full TYTAN operation (with Buffers) & 22{,}594 & 23.783\\
                \bottomrule
            \end{tabular}
            
            \begin{minipage}{\textwidth}
                \begin{tablenotes}
                    \item[*] \hspace{0.5cm}*\textit{Calculated at 950MHz clock frequency (1.053 ns per clock cycle)}
                \end{tablenotes}
            \end{minipage}
        \end{center}
        \label{tab3}
    \end{table}

\subsubsection{TYTAN Operation}
    TYTAN efficiently computes activation functions using a multiply-accumulate (MAC) unit modified for polynomial approximation. Input data enters through a dedicated buffer, while Taylor series coefficients are stored in TYTAN's internal buffer and reused for each input element.
        
    The unit iteratively computes polynomial approximation through successive multiplications and additions (Eq. 3). The first coefficient is added to zero and passed to the MAC unit. Subsequent operations multiply the accumulated result with input and add the next coefficient, enabling efficient evaluation of non-linear functions.
        
    Fig. \ref{fig:fsm} shows the finite state machine governing TYTAN operation. The FSM controls sequential processing through the computational pipeline, ensuring proper coefficient selection and accumulation at each polynomial evaluation step.
        
    TYTAN output is mapped to activation functions such as SELU, Sigmoid, Swish, GELU, Tanh and Softplus, each expressed in terms of MAC operations (Eqs. \ref{eq:4} to \ref{eq:9}). The architecture allows coefficient configuration through a dedicated port, providing flexibility to approximate diverse activation functions while maintaining scalability.
    
\subsubsection{Simulation and Synthesis}
    TYTAN is implemented in SystemVerilog HDL, with Taylor-series coefficients for the approximated activation functions derived from the software framework. A comprehensive verification testbench was developed using SystemVerilog to validate the functional correctness of TYTAN itself as well as with the SeLu, Sigmoid, and tanh add-ons. The testbench also counts the number of clock cycles it takes to finish the task across various modes of operations. A TYTAN is then simulated using both Synopsys VCS and Verilator, followed by synthesis using Synopsys Design Compiler V-2023.12-SP3 with the FreePDK45 process technology.
    
\subsection{Performance Evaluation}

\begin{figure*}[!t]
        \centering
        \subfigure[GeLu Function]{
            \includegraphics[width=2.0in]{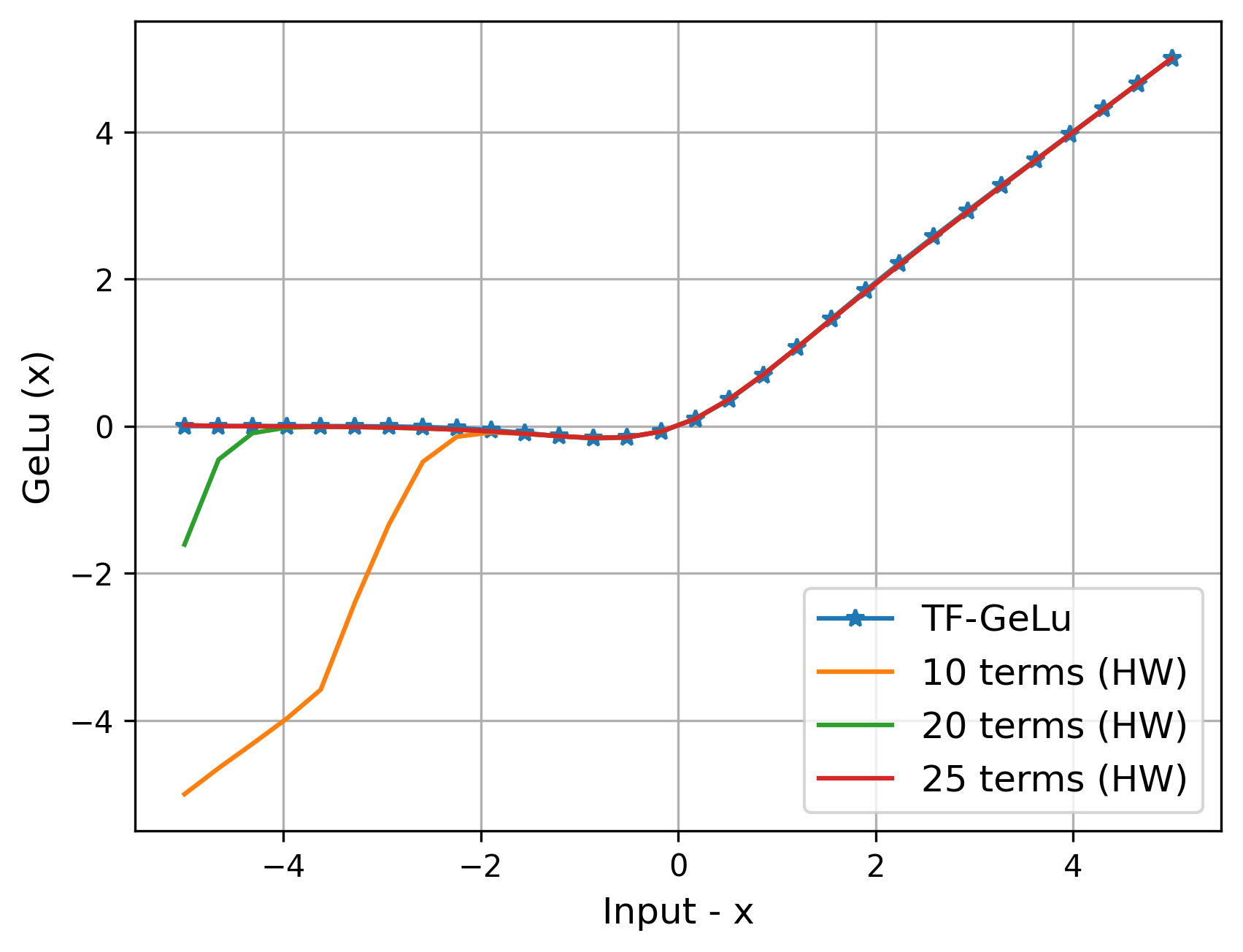}
            \label{fig:first_sub}
        }
        \subfigure[Sigmoid Function]{
            \includegraphics[width=2.0in]{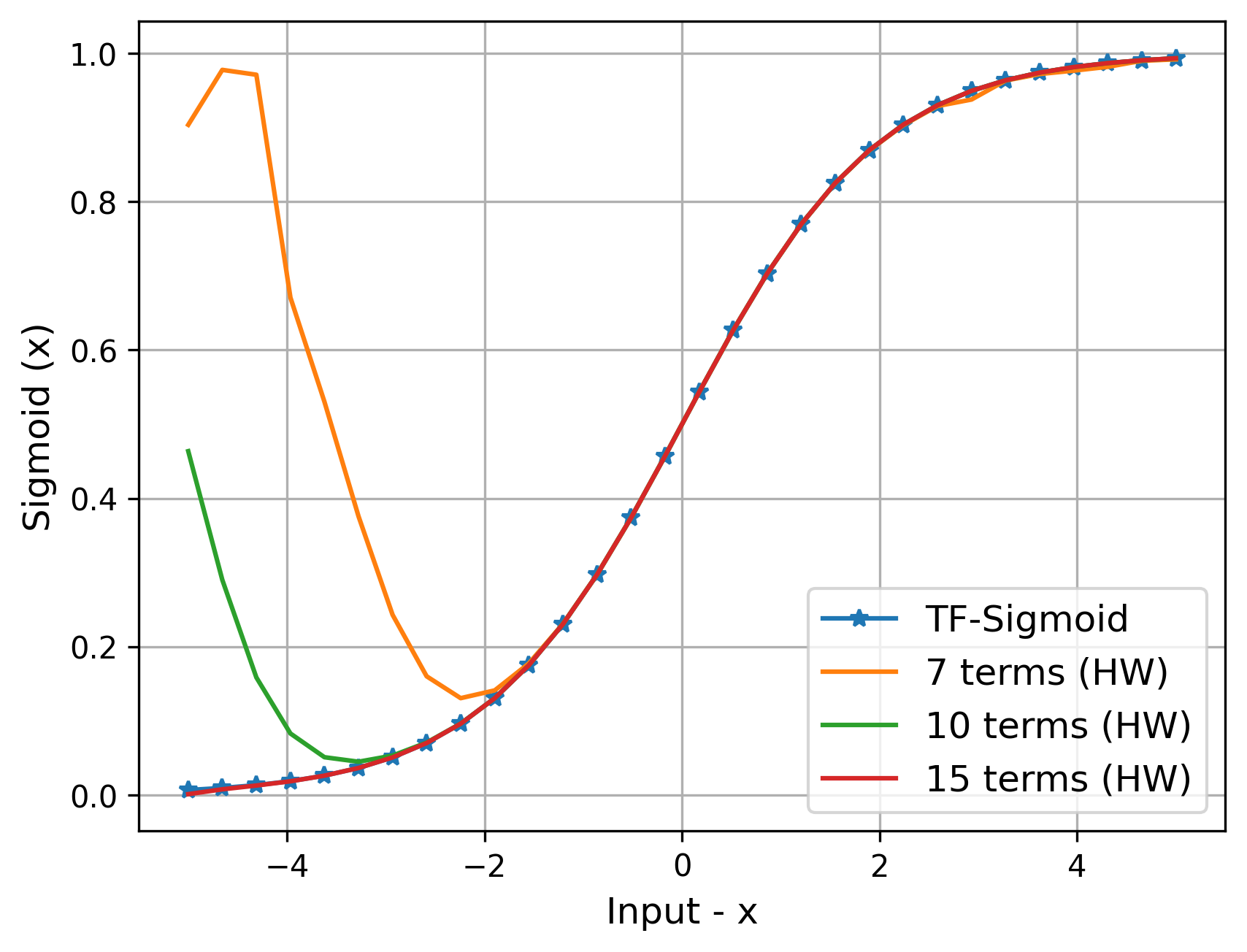}
            \label{fig:second_sub}
        }
        \\
        \subfigure[Swish Function]{
            \includegraphics[width=2.0in]{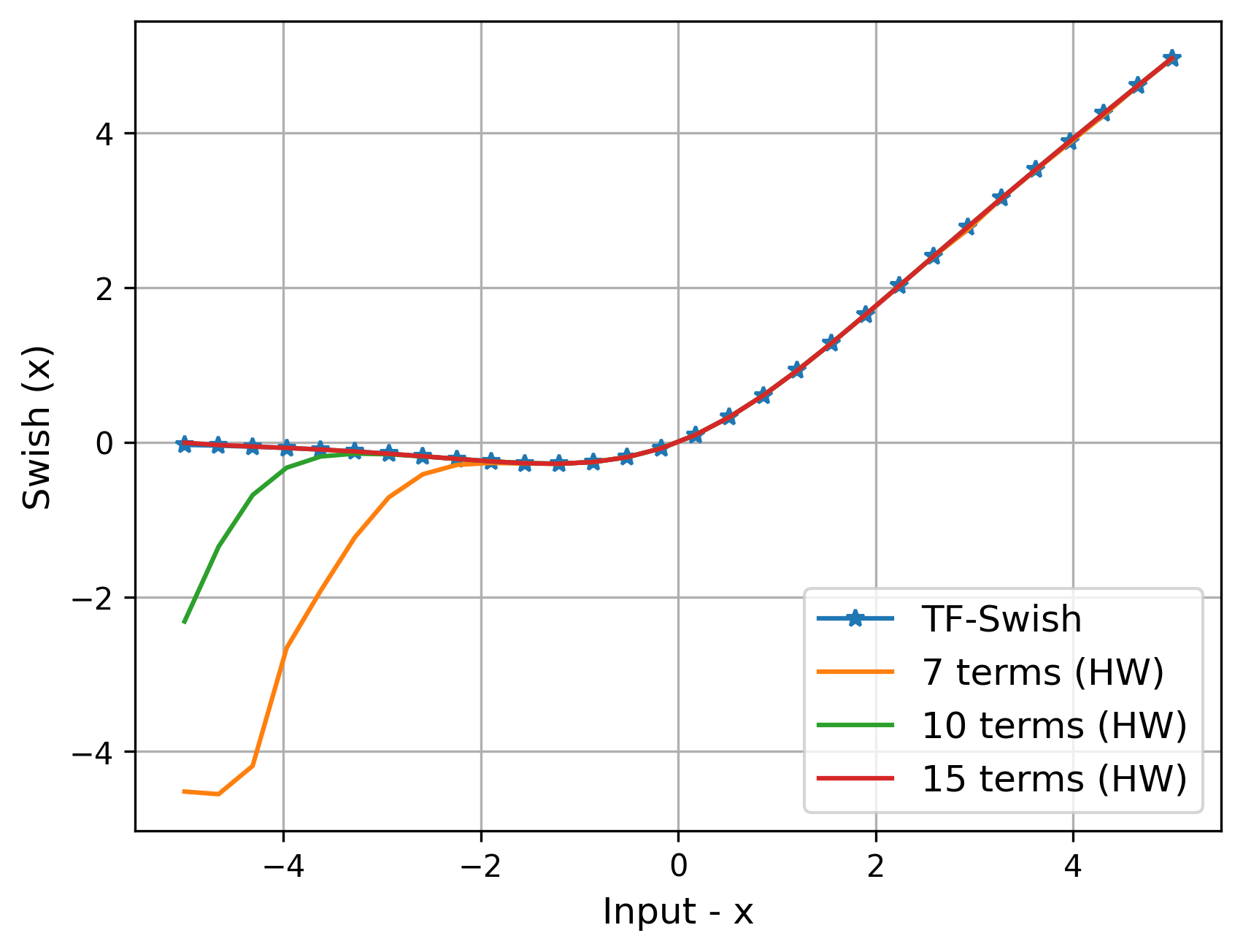}
            \label{fig:third_sub}
        }
        \subfigure[SeLu Function]{
            \includegraphics[width=2.0in]{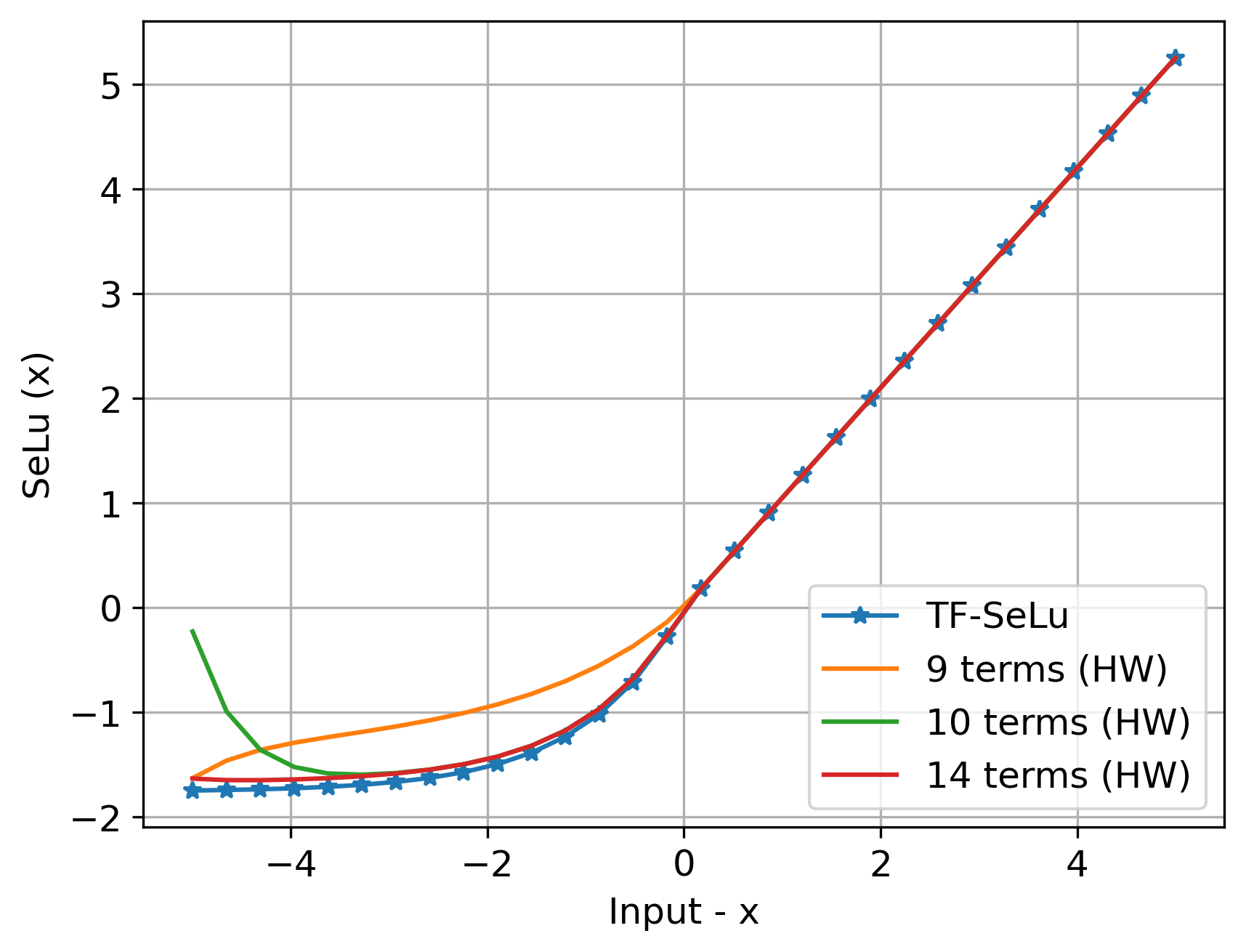}
            \label{fig:fourth_sub}
        }
        \\
        \subfigure[tanh Function]{
            \includegraphics[width=2.0in]{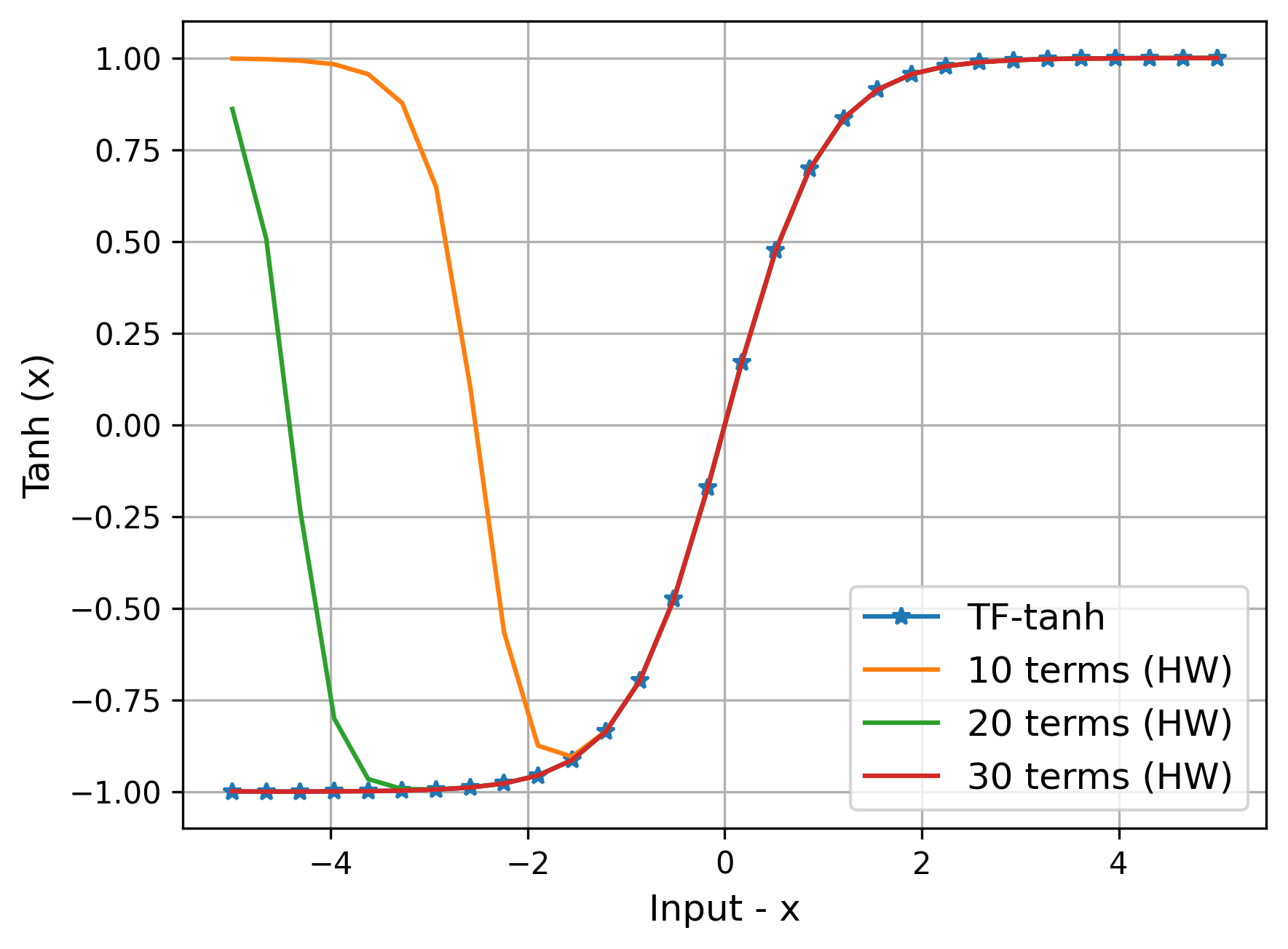}
            \label{fig:fifth_sub}
        }
        \subfigure[Softplus Function]{
            \includegraphics[width=2.0in]{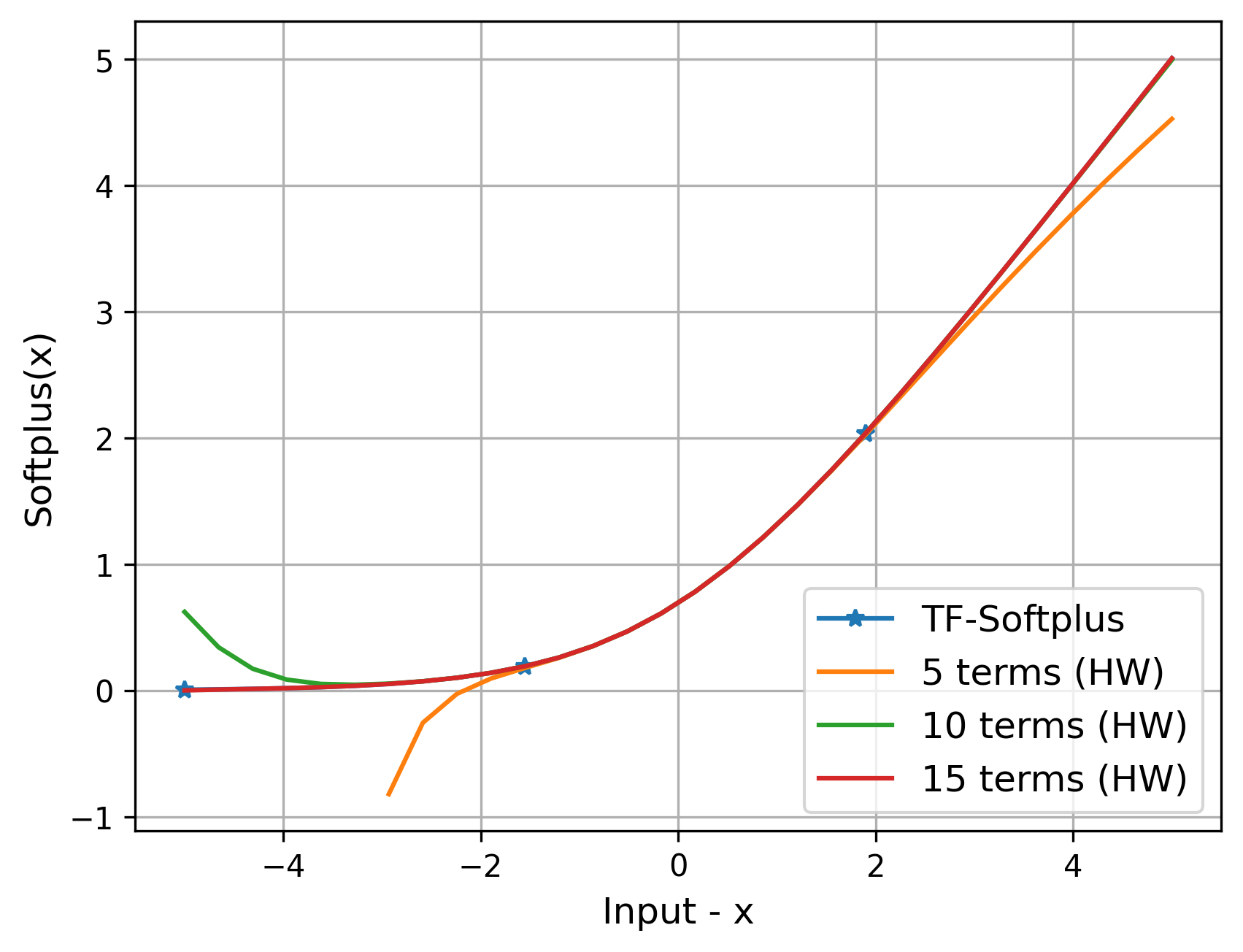}
            \label{fig:sixth_sub}
        }
        \caption{TensorFlow Activation Functions vs. Hardware-Approximated Activation Results}
        \label{fig:activation function}
    \end{figure*}

    Figure \ref{fig:activation function} presents a detailed comparison of the accelerator's accuracy across various activation functions, evaluated using different numbers of Taylor series terms. The results, benchmarked against TensorFlow's implementations over an input range from -5 to +5, reveal that increasing the number of Taylor series coefficients consistently enhances the hardware's accuracy. 

    Significantly, for every activation function, a threshold exists beyond which TYTAN's output precisely matches TensorFlow's reference. This finding implies that when the input range is limited to regions where the activation function is accurately represented with fewer coefficients, TYTAN can be optimally configured to use a reduced number of terms without sacrificing performance.

    The computational latency of TYTAN is determined exclusively by the number of Taylor series coefficients employed, irrespective of the specific activation function being computed. Table \ref{tab3} offers a comprehensive cycle analysis of the TYTAN hardware while calculating the tanh function using thirty Taylor series coefficients.

    The current implementation of TYTAN operates in FP32 precision, with the total latency comprising contributions from individual module delays including buffers, multiplier, and adder operations. This overall computational latency can be further reduced through strategic optimization approaches.

    By transitioning to lower precision formats or incorporating improved core hardware modules that is multiplier and adder circuits, the cumulative latency penalties from these core operations can be significantly minimized without changing rest of the system, thereby enhancing the accelerator's overall performance efficiency.

    \begin{table}[t]
        \centering
        \caption{Performance, Power, and Area (PPA) Comparison of SOTA Accelerators}
        \label{tab:comprehensive-ppa-comparison}
        \begin{tabular}{l>{\centering}p{2.2cm}>{\centering}p{2.2cm}>{\centering}p{2.2cm}>{\centering\arraybackslash}p{2.2cm}}
            \toprule
            \textbf{Design} & \textbf{Process Node} & \textbf{Area (mm$^2$)} & \textbf{Power (mW)} & \textbf{FMAX (MHz)} \\
            \midrule
            \textbf{TYTAN (This Work)} & 
              FreePDK45 &
              \begin{tabular}[t]{@{}c@{}}0.028 (base) \\ 0.037 (w/ NL)\end{tabular} &
              \begin{tabular}[t]{@{}c@{}}19.86 (base) \\ 24.37 (w/ NL)\end{tabular} &
              950 \\
            \addlinespace
            \textbf{NVDLA} & 
              FreePDK45 &
              1.002 &
              35.165 &
              450 \\
            \addlinespace
            \textbf{NN‑LUT~} & 
              7\,nm &
              0.001 &
              0.059 &
              N/A \\
            \addlinespace
            \textbf{UNO~} & 
              TSMC 45\,nm &
              0.283 (kernel) &
              66.5 (64 PE) &
              400 \\
            \addlinespace
            \textbf{ViTALiTy~} & 
              28\,nm CMOS &
              5.22 &
              1{,}460 &
              500 \\
            \bottomrule
        \end{tabular}
        
        \begin{minipage}{\textwidth}
            \begin{tablenotes}
                \item[] NL = Non-Linear addons; PE = Processing Elements
            \end{tablenotes}
        \end{minipage}
    \end{table}

\subsection{Comparative Analysis}
    Table \ref{tab:comprehensive-ppa-comparison} compares neural network accelerator designs, with NVDLA serving as the primary baseline for evaluating TYTAN's performance. The comparison focuses on NVDLA's p-partition that includes the Single Data Point Processor (SDP), which uses lookup tables for linear and non-linear functions on individual data points.

    TYTAN demonstrates exceptional performance improvements over NVDLA across all metrics using FreePDK45 technology. TYTAN achieves remarkable area efficiency with 35.785x improvement without addons and 27.081x with addons. Power consumption is reduced by 1.771x and 1.442x respectively, while maximum operating frequency more than doubles at 2.11x improvement. These results highlight TYTAN's scalability and modularity, efficiently handling additional capabilities like SeLu, Sigmoid, and tanh activation functions while maintaining substantial performance advantages.
    
    The table includes other state-of-the-art accelerators for context. NN-LUT uses advanced 7nm technology with extremely low area and power but requires 2 clock cycle latency and lacks frequency specifications, while TYTAN achieves 950 MHz in FreePDK45. UNO, despite using TSMC 45nm, consumes more power (66.5 mW vs 19.86-24.37 mW) and operates at lower frequency (400 MHz vs 950 MHz). ViTALiTy, implemented in more advanced 28nm CMOS, exhibits significantly higher area (5.22 mm²) and power (1,460 mW) while operating at lower frequency (500 MHz) than TYTAN's FreePDK45 implementation. These comparisons demonstrate TYTAN's exceptional efficiency, achieving superior performance even in a less advanced process technology.

    \begin{table}[t]
      \centering
      \caption{Precision and Operation Support Comparison}
      \label{tab:technical-specs}
      \begin{tabular}{lp{3.5cm}p{4.5cm}}
        \toprule
        \textbf{Design} & \textbf{Precision Support} & \textbf{Supported Operations} \\
        \midrule
        \textbf{TYTAN (This Work)} & 
          FP32 (configurable) &
          SELU, Sigmoid, Swish, GELU, Tanh, Softplus, etc \\
        \addlinespace
        \textbf{NVDLA (baseline)} & 
          INT8, INT16, FP16 &
          ReLU, PReLU, Sigmoid, Tanh \\
        \addlinespace
        \textbf{NN‑LUT~} & 
          FP32, FP16, INT32 &
          GELU, Softmax, LayerNorm \\
        \addlinespace
        \textbf{UNO~} & 
          16-bit fixed-point &
          Division, exp, log, sigmoid, tanh \\
        \addlinespace
        \textbf{ViTALiTy~} & 
          16-bit &
          Linear attention schemes \\
        \bottomrule
      \end{tabular}
    \end{table}


\section{Conclusion}
    This paper presents TYTAN, a hardware-software co-design for energy-efficient non-linear activation function computation in AI architectures. TYTAN achieves $\sim2.11\times$ frequency improvement, $\sim1.77\times$ power reduction, and $\sim35.8\times$ area reduction compared to NVDLA baseline, while outperforming other state-of-the-art accelerators even when implemented in 45nm technology.
    
    The modular and reconfigurable design ensures seamless integration with DNN accelerators, offering scalability from edge devices to high-performance systems. The parameterizable coefficient buffer optimizes precision-resource trade-offs, while supporting diverse AI workloads with configurable precision and a variety of activation functions. This approach delivers memory-optimized, accelerated activation computation with minimal resource constraints.
    The complete implementation is made available as open source \cite{tytan2025} to enable reproducibility and foster further research in this domain.

\bibliographystyle{splncs04}
\bibliography{VDAT}

\end{document}